\documentclass[%
reprint,
superscriptaddress,
 amsmath,amssymb,
 aps,
]{revtex4-2}

\usepackage{graphicx}
\usepackage{dcolumn}
\usepackage{bm}

\usepackage[caption=false]{subfig}
\usepackage{siunitx, soul, xcolor}
\usepackage{adjustbox}
\usepackage{gensymb}

\usepackage{xparse,xcoffins}

\ExplSyntaxOn
\NewCoffin\imagecoffin
\NewCoffin\labelcoffin

\keys_define:nn { miguel/label }
 {
  label   .tl_set:N = \l_miguel_label_tl,
  labelbox .bool_set:N = \l_miguel_label_box_bool,
  labelbox .default:n = true,
  fontsize .tl_set:N = \l_miguel_label_size_tl,
  fontsize .initial:n = \footnotesize,
  pos .choice:,
  pos/nw .code:n = \tl_set:Nn \l_miguel_label_pos_tl { left,up },
  pos/ne .code:n = \tl_set:Nn \l_miguel_label_pos_tl { right,up },
  pos/sw .code:n = \tl_set:Nn \l_miguel_label_pos_tl { left,down },
  pos/se .code:n = \tl_set:Nn \l_miguel_label_pos_tl { right,down },
  pos/n .code:n = \tl_set:Nn \l_miguel_label_pos_tl { hc,up },
  pos/w .code:n = \tl_set:Nn \l_miguel_label_pos_tl { left,vc },
  pos/s .code:n = \tl_set:Nn \l_miguel_label_pos_tl { hc,down },
  pos/e .code:n = \tl_set:Nn \l_miguel_label_pos_tl { right,vc },
  pos .initial:n = nw,
  unknown .code:n   = \clist_put_right:Nx \l_miguel_label_clist
                       { \l_keys_key_tl = \exp_not:n { #1 } }
 }
\clist_new:N \l_miguel_label_clist
\box_new:N \l_miguel_label_box
\box_new:N \l_miguel_label_image_box

\NewDocumentCommand{\xincludegraphics}{O{}m}
 {
  \group_begin:
  \tl_clear:N \l_miguel_label_tl
  \clist_clear:N \l_miguel_label_clist
  \keys_set:nn { miguel/label } { #1 }
  \tl_if_empty:NTF \l_miguel_label_tl
   {
    \miguel_includegraphics:Vn \l_miguel_label_clist { #2 }
   }
   {
    \SetHorizontalCoffin\imagecoffin
     {
      \miguel_includegraphics:Vn \l_miguel_label_clist { #2 }
     }
    \SetHorizontalCoffin\labelcoffin
     {
      \raisebox{\depth}
       {
        \bool_if:NTF \l_miguel_label_box_bool
         { \fcolorbox{white}{white}{\l_miguel_label_size_tl\l_miguel_label_tl} }
         { \l_miguel_label_size_tl\l_miguel_label_tl }
       }
     }
    \SetVerticalPole\imagecoffin{left}{-2pt+\CoffinWidth\labelcoffin/2}
    \SetVerticalPole\imagecoffin{right}{\Width-3pt-\CoffinWidth\labelcoffin/2}
    \SetHorizontalPole\imagecoffin{up}{\Height+5pt-\CoffinHeight\labelcoffin/2}
    \SetHorizontalPole\imagecoffin{down}{3pt+\CoffinHeight\labelcoffin/2}
    \use:x{\JoinCoffins\imagecoffin[\l_miguel_label_pos_tl]\labelcoffin[vc,hc]} 
    \TypesetCoffin\imagecoffin
   }
   \group_end:
 }
\NewDocumentCommand{\setlabel}{m}
 {
  \keys_set:nn { miguel/label } { #1 }
 }

\cs_new_protected:Nn \miguel_includegraphics:nn
 {
  \includegraphics[#1]{#2}
 }
\cs_generate_variant:Nn \miguel_includegraphics:nn { V }

\ExplSyntaxOff

\begin{document}
\preprint{APS/123-QED}

\title{Tuning the Curie temperature of a 2D magnet/topological insulator heterostructure to above room temperature by epitaxial growth}

\author{Wenyi Zhou}
\affiliation{The Ohio State University, Department of Physics, Columbus, OH, USA}
\author{Alexander J. Bishop}
\affiliation{The Ohio State University, Department of Physics, Columbus, OH, USA}
\author{Xiyue S. Zhang}
\affiliation{Cornell University, Department of Physics, Ithaca, NY, USA}
\author{Katherine Robinson}
\affiliation{The Ohio State University, Department of Physics, Columbus, OH, USA}\author{Igor Lyalin}
\author{Ziling Li}
\affiliation{The Ohio State University, Department of Physics, Columbus, OH, USA}
\author{Ryan Bailey-Crandell}
\affiliation{The Ohio State University, Department of Physics, Columbus, OH, USA}
\author{Thow Min Jerald Cham}
\affiliation{Cornell University, Department of Physics, Ithaca, NY, USA}
\author{Shuyu Cheng}
\affiliation{The Ohio State University, Department of Physics, Columbus, OH, USA}
\author{Yunqiu Kelly Luo}
\affiliation{Cornell University, Department of Physics, Ithaca, NY, USA}
\affiliation{Kavli Institute at Cornell, Ithaca, NY, USA}
\affiliation{University of Southern California, Department of Physics and Astronomy, Los Angeles, CA, USA}
\author{Daniel C. Ralph}
\affiliation{Cornell University, Department of Physics, Ithaca, NY, USA}
\affiliation{Kavli Institute at Cornell, Ithaca, NY, USA}
\author{David A. Muller}
\affiliation{Cornell University, Department of Physics, Ithaca, NY, USA}
\affiliation{Kavli Institute at Cornell, Ithaca, NY, USA}
\author{Roland K. Kawakami}
\email[email address: ]{kawakami.15@osu.edu}
\affiliation{The Ohio State University, Department of Physics, Columbus, OH, USA}

\date{\today} 

\begin{abstract}
Heterostructures of two-dimensional (2D) van der Waals (vdW) magnets and topological insulators (TI) are of substantial interest as candidate materials for efficient spin-torque switching, quantum anomalous Hall effect, and chiral spin textures. 
However, since many of the vdW magnets have Curie temperatures below room temperature, we want to understand how materials can be modified to stabilize their magnetic ordering to higher temperatures. 
In this work, we utilize molecular beam epitaxy to systematically tune the Curie temperature ($T_C$) in thin film Fe$_3$GeTe$_2$/Bi$_2$Te$_3$ from bulk-like values ($\sim$220 K) to above room temperature by increasing the growth temperature from 300 $\degree$C to 375 $\degree$C.
For samples grown at 375 $\degree$C, cross-sectional scanning transmission electron microscopy (STEM) reveals the spontaneous formation of different Fe$_m$Ge$_n$Te$_2$ compositions (e.g. Fe$_5$Ge$_2$Te$_2$ and Fe$_7$Ge$_6$Te$_2$) as well as intercalation in the vdW gaps, which are possible origins of the enhanced Curie temperature. This observation paves the way for developing various Fe$_m$Ge$_n$Te$_2$/TI heterostructures with novel properties.

\end{abstract}


\maketitle

\section{Introduction} \label{sec:intro}

Van der Waals (vdW) materials have been explored intensely due to many exotic phenomena arising in two-dimensional (2D) sheets and heterostructures that can be tuned by electrostatic gating, strain, and stacking orientation \cite{liu_van_2016, novoselov_2d_2016, wang_magnetic_2022, gibertini_magnetic_2019, li_emergent_2022}. Recently, there has been tremendous interest in magnetic atomically-thin crystals (2D magnets) and their integration with topological insulators (TI) \cite{hou_magnetizing_2019, mogi_large_2019, wang_above_2020, mogi_current-induced_2021, liu_magnetic_2021}. These are attractive candidate materials for novel topological states and highly efficient spin-orbit torque (SOT) magnetic switching \cite{mellnik_spin-transfer_2014}, and an important challenge is to create epitaxial 2D magnet/TI heterostructures with sharp interfaces and enhanced Curie temperatures ($T_C$).

For novel topological states, the long-sought quantum anomalous Hall effect (QAHE) was realized in related vdW materials. In 2013, the QAHE was first observed in a magnetically-doped topological insulator, specifically Cr-doped (Bi,Sb)$_2$Te$_3$, at low temperatures ($<100$ mK) \cite{chang_experimental_2013} and later in the intrinsic magnetic topological insulator MnBi$_2$Te$_4$ at higher temperatures (5 – 10 K) \cite{deng_quantum_2020}. The 2D magnet/TI heterostructures are among the family of magnetic topological insulators that may generate QAHE at further elevated temperatures.\cite{liu_magnetic_2021}
For efficient SOT switching, 2D magnet/TI heterostructures are attractive due to the large spin-torque efficiency of TIs and the small magnetic volume of 2D magnets. But to utilize these for energy-efficient magnetic memory and logic, it will be crucial to synthesize epitaxial heterostructures for scalability while maintaining high-quality interfaces to preserve the topological surface states and having $T_C$ above room temperature.

Achieving these properties simultaneously has been challenging. Our initial efforts to integrate room temperature 2D magnet MnSe$_2$ \cite{ohara_room_2018} with TI Bi$_2$Se$_3$ produced interdiffusion of Mn into the Bi$_2$Se$_3$ \cite{noesges_chemical_2020} ultimately leading to the formation of MnBi$_2$Se$_4$, a magnetic topological insulator with layered antiferromagnetic order at low temperatures ($<$ 10 K) \cite{zhu_synthesis_2021}. 
Alternatively, a promising material is Fe$_3$GeTe$_2$ (atomic structure shown in Fig.~\ref{fig:FGT_on_BT_atomic}), which has a relatively high $T_C$ of $\sim 220$ K for bulk crystals and a strong perpendicular magnetic anisotropy (PMA).
In addition to Fe$_3$GeTe$_2$, there have been a few reports of ferromagnets with higher $T_C$ among the Fe$_m$Ge$_n$Te$_2$ (FGT) family. Bulk Fe$_5$Ge$_2$Te$_2$ was synthesized and characterized to have large PMA up to $\sim 250$ K \cite{jothi_fe_2020}, while bulk Fe$_5$GeTe$_2$ has a $T_C$ of $\sim 310$ K \cite{may_ferromagnetism_2019}. 
Bulk Fe$_4$GeTe$_2$ was reported to show a $T_C$ of $\sim 270$ K \cite{seo_nearly_2020}. 
Recently, 4\,nm Fe$_4$GeTe$_2$ films grown on sapphire by MBE were reported to show a $T_C$ of $\sim 530$ K, with tunable magnetic anisotropy by slightly varying the Fe concentration \cite{wang_interfacial_2023}.
Also, epitaxial growth of Fe$_5$GeTe$_2$ with weak PMA and a $T_C$ of $\sim 293$ K was achieved by MBE \cite{ribeiro_large-scale_2022}. Furthermore, $T_C$ above room temperature has been observed for Fe$_3$GeTe$_2$ in a variety of contexts including electrostatic gating using ionic liquids \cite{deng_gate-tunable_2018}, patterning-induced enhancement of $T_C$ \cite{li_patterning-induced_2018}, and through ion intercalation in which case a large variation in the density of states (DOS) at the Fermi level due to extreme electron doping induced by the ionic gate leads to modulation in the ferromagnetism \cite{li_electrochemical_2021}. Deposition of Fe$_3$GeTe$_2$ onto Bi$_2$Te$_3$ has also been reported to help elevate $T_C$ of Fe$_3$GeTe$_2$ \cite{wang_above_2020}. However, the origins of the enhanced $T_C$ remain unclear.

To address the important issue of $T_C$ enhancement, it is essential to systematically investigate the relationship between the magnetic properties and atomic-scale structure of FGT and FGT/TI heterostructures. To this end, molecular beam epitaxy (MBE) and scanning transmission electron microscopy (STEM) are ideal complementary tools that respectively provide atomically-precise layer-by-layer deposition and cross-sectional imaging with atomic resolution.
Furthermore, several groups have successfully applied MBE to synthesize Fe$_3$GeTe$_2$ films \cite{liu_wafer-scale_2017, roemer_robust_2020, wang_above_2020, zhou_kinetically_2022}.

In this Letter, we utilize MBE and STEM to demonstrate the systematic tuning of $T_C$ in Fe$_3$GeTe$_2$/Bi$_2$Te$_3$ from bulk-like values ($\sim$220 K) to above room temperature by varying the growth conditions. 
Specifically, we investigate the growth of Fe$_3$GeTe$_2$ films onto Bi$_2$Te$_3$ with growth temperatures from 300 $\degree$C to 375 $\degree$C. Using reflection high-energy electron diffraction (RHEED) during growth, we observe streaky patterns that indicate high-quality epitaxial growth with well-defined in-plane orientation for all growth temperatures.
Through variable-temperature magneto-optic measurements, we find that samples grown at 300\,$\degree$C exhibit a $T_C$ of $\sim 220$ K, similar to bulk Fe$_3$GeTe$_2$. Increasing the growth temperature systematically enhances $T_C$, and the $T_C$ exceeds 320 K for samples grown at 375\,$\degree$C.
Origins of the $T_C$ enhancement are identified through surprising observations by cross-sectional STEM.
While growth at 325 $\degree$C produces uniform Fe$_3$GeTe$_2$ on Bi$_2$Te$_3$, STEM images of samples grown at 375 $\degree$C reveal the
spontaneous formation of different Fe$_m$Ge$_n$Te$_2$ compositions
including those with thicker vdW layers (e.g. Fe$_5$Ge$_2$Te$_2$ and Fe$_7$Ge$_6$Te$_2$) and evidence of intercalation in the van der Waals gap. 
This signifies the onset of a new growth window that supports the formation of multiple members of the FGT family, and importantly, the interface with Bi$_2$Te$_3$ remains very sharp. These observations constitute a major advance enabling the future integration of a broader family of FGT materials (Fe$_m$Ge$_n$Te$_2$) with topological insulators, while keeping the high quality of both the 2D magnet and TI layers.

\section{Epitaxial growth and structural characterization} \label{sec:growth}


\begin{figure}
\setlabel{pos=nw,fontsize=\scriptsize} 
  \subfloat{\xincludegraphics[width=0.45\textwidth,label=(a)]{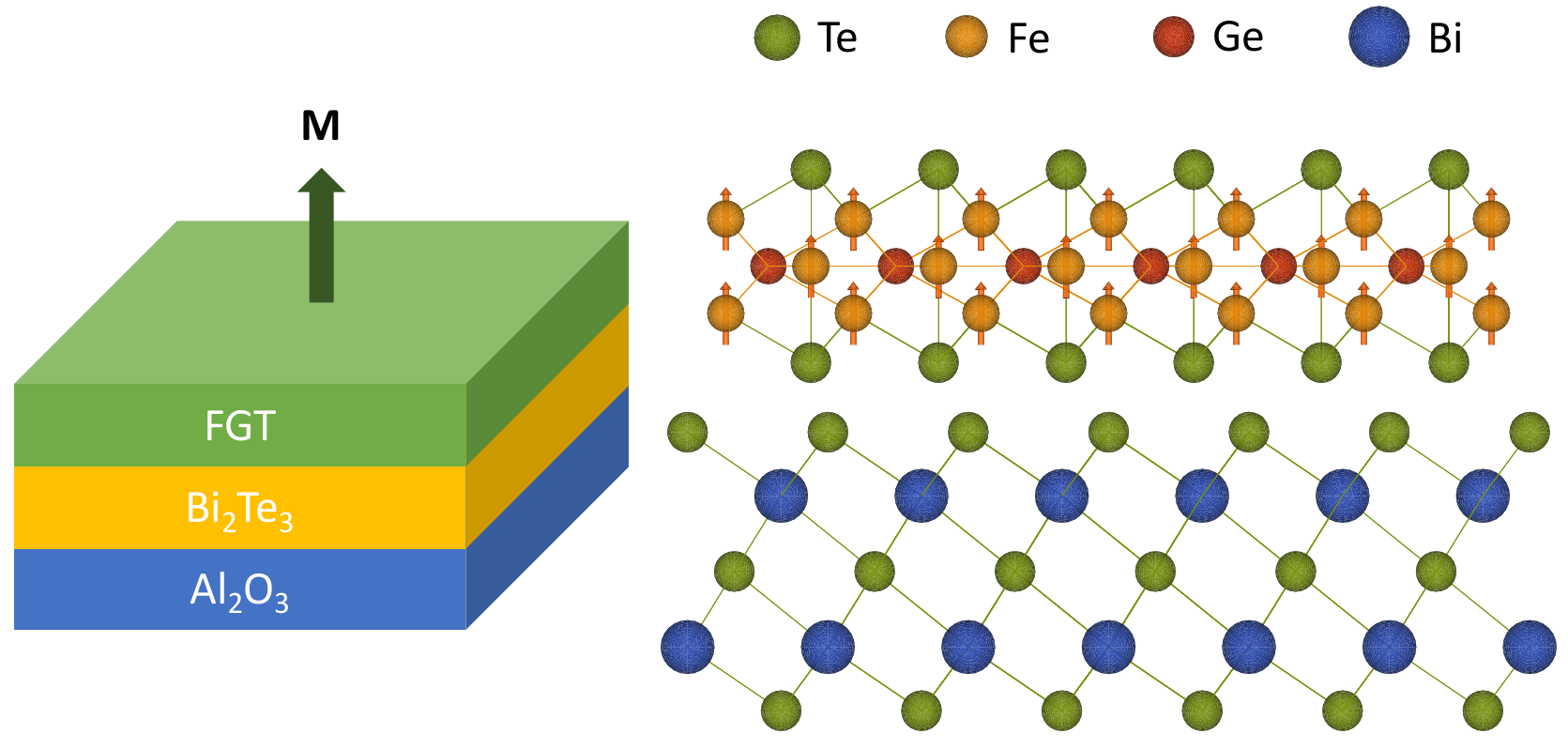}\label{fig:FGT_on_BT_atomic}}
  \hfill
  \setlabel{pos=se,fontsize=\scriptsize}
  \subfloat{\xincludegraphics[width=0.23\textwidth,label=\textcolor{white}{(b)}]{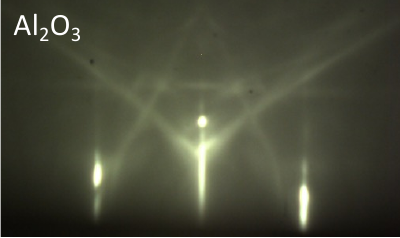}\label{fig:RHEED_sapphire}}
  \hfill
  \subfloat{\xincludegraphics[width=0.23\textwidth,label=\textcolor{white}{(c)}]{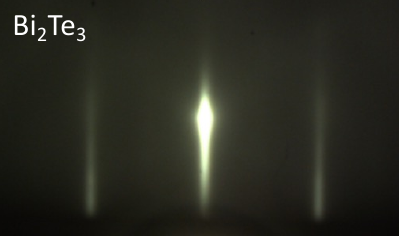}\label{fig:RHEED_BT}}
  \hfill
  \subfloat{\xincludegraphics[width=0.23\textwidth,label=\textcolor{white}{(d)}]{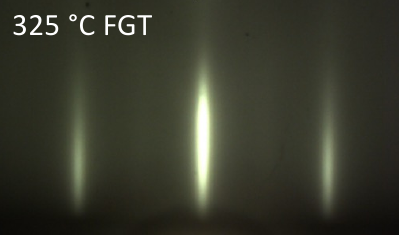}\label{fig:RHEED_325C}}
  \hfill
  \subfloat{\xincludegraphics[width=0.23\textwidth,label=\textcolor{white}{(e)}]{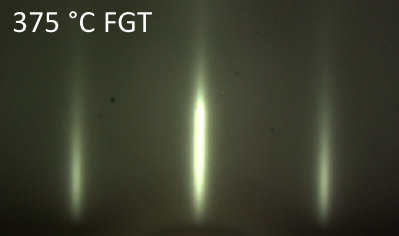}\label{fig:RHEED_375C}}
  \hfill

\caption{(a) Schematic of a FGT/Bi$_2$Te$_3$ heterostructure with out-of-plane magnetization and the atomic lattice of FGT and Bi$_2$Te$_3$. (b) RHEED pattern for an Al$_2$O$_3$(0001) substrate. (c) RHEED pattern for an 8\,nm film of Bi$_2$Te$_3$ grown on an Al$_2$O$_3$(0001) substrate. (d,e) RHEED patterns for 4\,nm films of FGT grown on Bi$_2$Te$_3$ at 325\,$\degree$C and 375\,$\degree$C, respectively. The electron beam is along [10$\bar{1}$0] direction, and the hexagonal unit cells of all layers and the substrate share the same in-plane alignment. \label{fig:RHEED}
}
\end{figure}

For all FGT films grown, we targeted the Fe$_3$GeTe$_2$ composition but found that the resulting material could be more complex. The FGT/Bi$_{2}$Te$_{3}$ films are deposited in a Veeco 930 MBE system with a base pressure of $2 \times 10^{-10}$ Torr.  The Al$_{2}$O$_{3}$(0001) substrates (MTI Corporation) are prepared by sonication in acetone, methanol and isopropyl alcohol for 5 min each before being annealed in air for 3 hours at 1000 \degree C. Then substrates are mounted on Omicron flag style paddles, held in a Veeco UNI-Block with paddle adapter, and loaded into the MBE chamber. After degassing at 800 \degree C for an hour, the Al$_{2}$O$_{3}$ substrates are cooled down to 255 \degree C for starting the growth of FGT/Bi$_{2}$Te$_{3}$ heterostructure films. For synthesis, we first grow the Bi$_{2}$Te$_{3}$ layer on Al$_{2}$O$_{3}$(0001) substrate by co-depositing Bi (99.95\%, Alfa Aesar) and Te (99.9999\%, United Mineral Corp.) with an atomic flux ratio of 1:15. The high-quality Bi$_{2}$Te$_{3}$ films are synthesized by a two-step growth method \cite{harrison_two-step_2013} with a lower temperature (255 \degree C) in the first step and a relatively higher temperature (295 \degree C) in the second step both with a growth rate of $\sim$0.03\,\AA\,/s. Then the FGT films on top of Bi$_{2}$Te$_{3}$ are grown by co-depositing Fe (99.99\%, Alfa Aesar), Ge (99.9999\%, Alfa Aesar), and Te with an atomic flux ratio of 3:1:20, as measured by a beam flux monitor (BFM) and calibrated by x-ray reflectivity (XRR) of elemental films. The growth rate for FGT films is $\sim$0.09\,\AA\,/s. We note that during all temperature ramping procedures, the film was always exposed to an overpressure of Te flux. At the end of the growth, samples are moved from the growth chamber to the buffer chamber to be cooled down relatively quickly from growth temperature to prevent possible structural change or Te deficiency of the film surface. The growth temperatures are read by a thermocouple behind the paddle and UNI-Block and in the center of the heater, so the actual temperature of the substrate could be lower than the thermocouple reading.
After growth, all the samples in this study are transferred via an ultrahigh-vacuum suitcase to another chamber where they are capped with 10 nm of CaF$_{2}$ to protect against oxidation.

We investigate the structural properties of the FGT/Bi$_2$Te$_3$ heterostructures using a wide range of characterization techniques. \textit{In situ} reflection high energy electron diffraction (RHEED) is employed to investigate the crystal structure and flatness of the surface during growth. After growth, we utilize x-ray diffraction (XRD) to probe the bulk crystal structure (see Supplementary Material \cite{SupplMat}, Section 1) and cross-sectional scanning transmission electron microscopy (STEM) to probe the atomic scale structure. 
The chemical composition is determined by electron energy loss spectroscopy (EELS) as follows. The material structure is first obtained by high-angle annular dark-field (HAADF) and bright field (BF) imaging, and EELS is then performed at each position so that the line cut of EELS is registered to the atomic structure. The EELS line cut along the growth direction ($z$) is obtained by integrating over the horizontal plane at specific $z$-positions within the specimen.

\begin{figure*}
\setlabel{pos=nw,fontsize=\scriptsize}
\setlabel{pos=nw,fontsize=\scriptsize} 
\subfloat{\xincludegraphics[width=0.37\textwidth,label={(a)}]{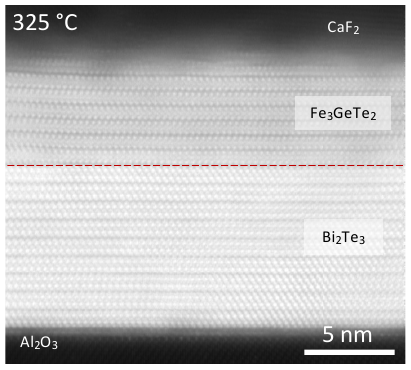}\label{fig:STEM325}}
\hfill
\subfloat{\xincludegraphics[width=0.61\textwidth,label={(b)}]{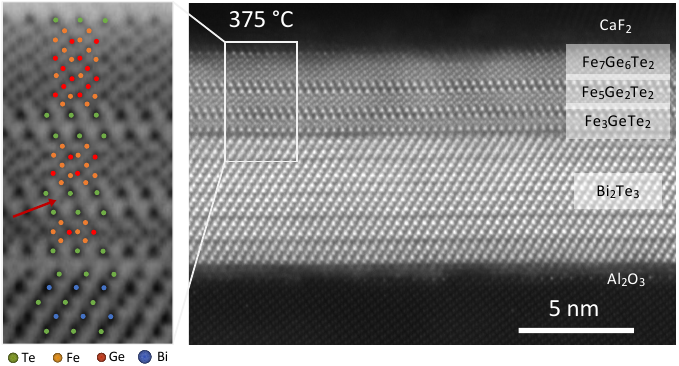}\label{fig:STEM375}}
\hfill
\subfloat{\xincludegraphics[width=0.443\textwidth,label={(c)}]{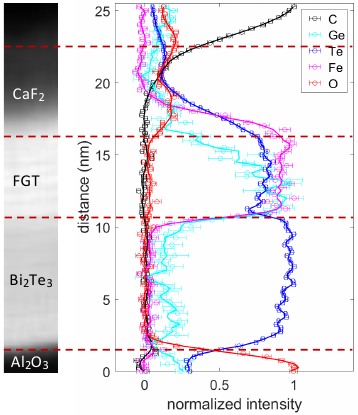}\label{fig:EELS325}}
\hfill
\subfloat{\xincludegraphics[width=0.457\textwidth,label={(d)}]{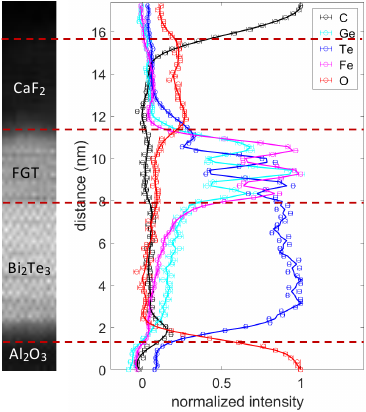}\label{fig:EELS375}}
\hfill

\caption{STEM investigations. (a) HAADF STEM image of an FGT(6.4\,nm)/Bi$_2$Te$_3$(10\,nm) heterostructure grown at 325\,$\degree$C (for FGT). (b) HAADF STEM image of an FGT(4\,nm)/Bi$_2$Te$_3$(8\,nm) heterostructure grown at 375\,$\degree$C (for FGT) and a zoomed in BF STEM image with different atoms marked out in the left. The red arrow indicates the position of an intercalant atom. (c) EELS line profile of the FGT(6.4\,nm)/Bi$_2$Te$_3$(10\,nm) heterostructure grown at 325\,$\degree$C (for FGT) with the corresponding HAADF image on the left. (d) EELS line profile of the FGT(4\,nm)/Bi$_2$Te$_3$(8\,nm) heterostructure grown at 375\,$\degree$C (for FGT) with the corresponding HAADF image on the left.  \label{fig:XRD_STEM}
}
\end{figure*}
 
We characterize the epitaxial growth using RHEED, which exhibits streaky patterns throughout the entire growth of the FGT/Bi$_2$Te$_3$/Al$_2$O$_3$(0001) heterostructure. Beginning with the Al$_2$O$_3$ substrate (Figure~\ref{fig:RHEED_sapphire}), the growth of 8 nm Bi$_2$Te$_3$ produces streaky patterns indicative of two-dimensional epitaxial growth (Figure~\ref{fig:RHEED_BT}). Following the deposition of 4 nm of FGT ($\sim 5$ vdW layers) onto the Bi$_2$Te$_3$ layer, we observe streaky patterns as shown in Figures \ref{fig:RHEED_325C} and \ref{fig:RHEED_375C} for FGT growth temperatures of 325\,$\degree$C and 375\,$\degree$C, respectively.
These patterns are characteristic of high-quality two-dimensional layer-by-layer growth with atomically-flat terraces of finite size.
The in-plane lattice constant of our FGT films extracted from RHEED patterns is $\sim 3.985$ \AA, which is very close to the reported values $a=b=3.991$ \AA\, for FGT in the literature \cite{deiseroth_fe3gete2_2006, yi_competing_2016, roemer_robust_2020}. 
In terms of crystallographic properties, the uniformly spaced streaks indicate that the FGT lattice remains aligned with the Bi$_2$Te$_3$ underlayer.
In addition, the streaky patterns are maintained throughout the growth, which suggests a high crystalline quality of the interface.
RHEED patterns for 300\,$\degree$C and 350\,$\degree$C growths are shown in the Supplementary Material~\cite{SupplMat} Section 2.

To investigate the crystal structure of the samples, we utilize HAADF-STEM for atomic-scale imaging of the cross-section.
Figure \ref{fig:STEM325} shows an image for a $\sim$ 6.4 nm ($\sim 8$\,vdW layers) FGT film grown at 325$\degree$C on Bi$_2$Te$_3$, acquired along the FGT [11$\bar{2}$0] direction. 
vdW gaps are distinguished clearly for both the FGT and Bi$_{2}$Te$_{3}$, indicating high-quality vdW layers. In addition, the interface between FGT and Bi$_{2}$Te$_{3}$ indicated by the dashed red line is very sharp and free of any secondary phases.

The right image of Fig. \ref{fig:STEM375} shows a HAADF image for an FGT/Bi$_{2}$Te$_{3}$ film intended to have $\sim$ 4\,nm of Fe$_3$GeTe$_2$ using a growth temperature of 375$\degree$C. vdW layers are again easily distinguished throughout the entire film, however, the FGT vdW layers appear to be not only Fe$_3$GeTe$_2$ layers but also compositions of thicker FGT layers, as opposed to the uniform thicknesses in the 325\degree C sample.
We identify at least three different compositions of FGT layers in the HAADF images. 
The left image of Fig.~\ref{fig:STEM375} is a magnified bright field (BF) image with the Fe, Ge, Bi, and Te atoms marked out by dots of different colors. 
Because the brightness of atoms in the BF (HAADF) image is sensitive to atomic number and heavier atoms appear to be darker (brighter), we can distinguish Fe, Ge, and Te from the contrast and make a reasonable estimation about the FGT composition. 
Based on such analysis of the BF and HAADF images, besides Fe$_3$GeTe$_2$, we can see the existence of thicker cousins like Fe$_5$Ge$_2$Te$_2$ and Fe$_7$Ge$_{6}$Te$_2$ are also stabilized simultaneously in our film at the temperature of 375\degree C. 
We also notice that there are intercalants in the vdW gaps (one is indicated by the red arrow), and we estimate they are Fe atoms based on the lower concentration of Fe in the thicker FGT layers.
The dependence of the Fe and Ge concentrations with depth (along the growth direction) is shown in the normalized EELS line profiles (Fig.~\ref{fig:EELS375}), which illustrates the variation of the chemical composition within the FGT layers.

The atomic structures of Bi$_{2}$Te$_{3}$, Fe$_3$GeTe$_2$ and Fe$_5$Ge$_2$Te$_2$ observed in our sample match very well with what has been reported in the literature. The out-of-plane lattice constants for Bi$_2$Te$_3$, Fe$_3$GeTe$_2$, and Fe$_5$Ge$_2$Te$_2$ are 30.450\,\AA~\cite{hosokawa_growth_2019}, 16.333\,\AA~\cite{deiseroth_fe3gete2_2006} and 10.778\,\AA~\cite{jothi_fe_2020} where the unit cell contains 3, 2, and 1 vdW layers respectively, leading to the thickness of the vdW layers to be 10.150\,\AA, 8.167\,\AA\, and 10.778\,\AA. In our samples, the Bi$_2$Te$_3$ and Fe$_3$GeTe$_2$ vdW layers have thicknesses of 10.492\,\AA\, and 8.290\,\AA, which agree with the literature values of 10.150\,\AA\, and 8.167\,\AA, respectively \cite{hosokawa_growth_2019, deiseroth_fe3gete2_2006}.
The vdW monolayer of Fe$_5$Ge$_2$Te$_2$ indicates a thickness of $\sim$11.183\,\AA, which is close to the literature value 10.778\,\AA~\cite{jothi_fe_2020}.
On the other hand, the top Fe$_7$Ge$_6$Te$_2$ vdW layer exhibits a thickness of $\sim$16.685\,\AA, and to our knowledge, such films have not been observed previously in experiments.

\begin{figure*}[t]
  \adjustbox{minipage=1.1em,valign=b}
  \centering
    \includegraphics[width=1.0\textwidth]{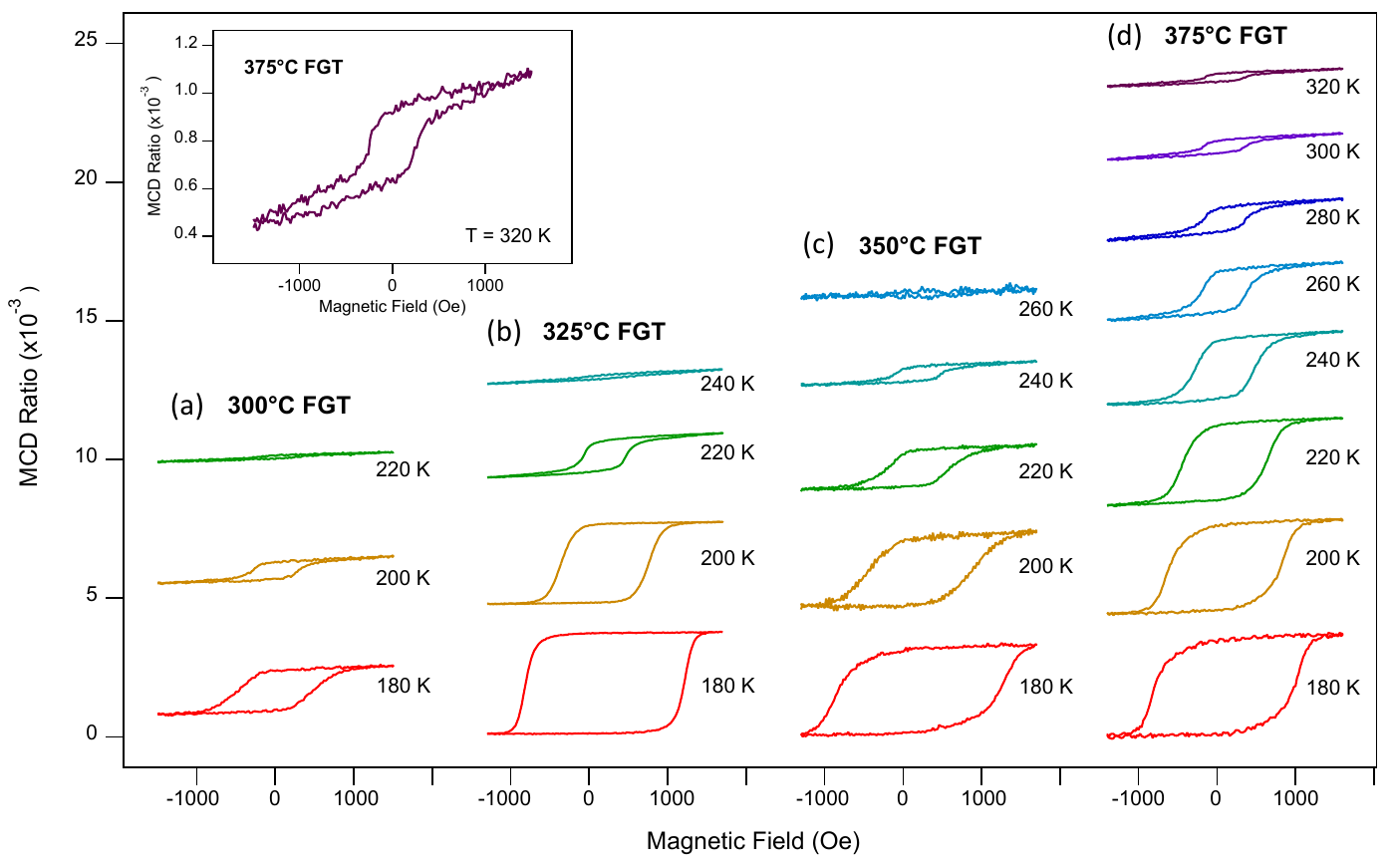}
\hfill
\caption{Temperature dependence of MCD loops for FGT samples ($\sim$ 4\,nm) grown at
300\,$\degree$C, 325\,$\degree$C, 350\,$\degree$C, and 375\,$\degree$C (left to right) on Bi$_2$Te$_3$. The Curie temperature exhibits a systematic increase with increasing growth temperature. Inset: zoom-in of the highest temperature ferromagnetic loop, which was observed for the 375\,$\degree$C sample at 320\,K (maximum temperature of the MCD system).\label{fig:MCDloops}
}
\end{figure*}

\begin{figure}
  \adjustbox{minipage=1.1em,valign=b}
  \centering
 \includegraphics[width=0.48\textwidth]{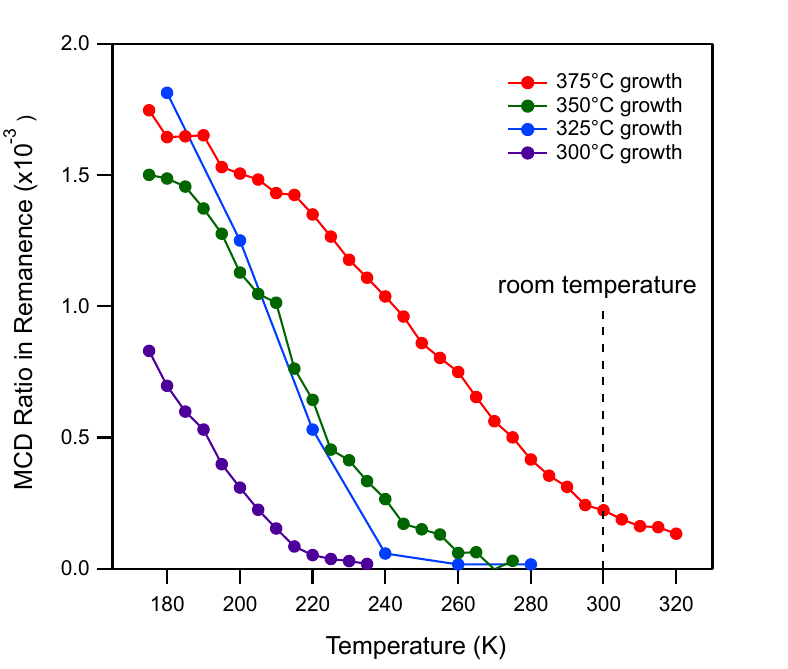}
\hfill
\caption{Summary of magnetization vs temperature. Calculated MCD ratio in remanence at different temperatures for 300\,$\degree$C, 325\,$\degree$C, 350\,$\degree$C, and 375\,$\degree$C grown FGT/Bi$_2$Te$_3$ samples. The increase of Curie temperature to above room temperature is clearly shown.}\label{fig:MvsT}
\end{figure}

\section{Magnetic properties} \label{sec:properties}
The magneto-optical technique of reflective magnetic circular dichroism (MCD) enables us to measure out-of-plane magnetic hysteresis loops of the FGT/Bi$_2$Te$_3$ heterostructures. A continuous wave 532 nm laser is focused onto the sample with a $45\degree$ angle of incidence and the reflected beam intensity ($I$) is measured using a photodiode detector (Thorlabs PDA36A2). The polarization of the incident beam is modulated between right-circular polarized (RCP) and left-circular polarized (LCP) at a frequency of 50 kHz using a photoelastic modulator (PEM) (Hinds Instruments) and the MCD ratio $(\frac{I_{RCP}-I_{LCP}}{I_{RCP}+I_{LCP}})$ is measured using a lock-in amplifier (Signal Recovery 7270). We apply a magnetic field along the surface normal to measure out-of-plane magnetic hysteresis loops and hold the sample under vacuum in an optical cryostat to vary the temperature. The typical magnetic field ramp rate for the hysteresis loop is 1000 Oe/s and the presented loop is an average over several (typically 5-20) scans.
The fast field ramp rate is enabled by using a data acquisition card (National Instruments PCIe-6323) to control the electromagnet current supply (Kepco BOP) via analog output voltage and to read the lock-in amplifier via analog input voltage. 
A typical acquisition rate is 20 ms per data point with a lock-in time constant of 5 ms, which provides lock-in averaging over $\sim$250 cycles (50 kHz $\times$ 5 ms).
The fast acquisition helps avoid drift of the MCD signal.


Figure \ref{fig:MCDloops} shows representative hysteresis loops measured on 4\,nm FGT samples grown on Bi$_2$Te$_3$ at $300 \degree$C (a), $325 \degree$C (b), $350 \degree$C (c), and $375 \degree$C (d).
All four samples have some common characteristics. 
They exhibit perpendicular magnetic anisotropy with square hysteresis loops of nearly 100\% remanence at low temperatures.
The coercivity increases as the measurement temperature is lowered, and we are unable to capture full hysteresis loops at temperatures below those shown in Figure \ref{fig:MCDloops} due to the limited range of magnetic field of the measurement system ($\sim1500$ Oe).

One of the most interesting aspects of the data is the variation of $T_C$ with growth temperature. In all samples, the magnitude of the MCD decreases as the measurement temperature is increased. 
The trend of $T_C$ enhancement with increasing growth temperature is visible from the hysteresis loops. 
For the 300 $\degree$C sample, the remanence appears to be smaller and disappears at $\sim 220$\,K, which is consistent with the bulk $T_C$ of FGT.
However, at the same 220\,K temperature, the 325 $\degree$C grown sample shows a very clear opening in the hysteresis loop, which persists up to $\sim 240$\,K.
In the meantime, we notice that the shape of the hysteresis loops for the 325 $\degree$C sample is more square compared to the others. Similarly, $T_C$ continues to increase with higher growth temperatures 350 $\degree$C and 375 $\degree$C. 
Interestingly, the remanence is still observed to be present at 320 K for the $375 \degree$C sample, which is zoomed in and shown in the inset of Figure \ref{fig:MCDloops}.

In Figure \ref{fig:MvsT}, we provide a better illustration of the temperature dependence of the magnetization by plotting the magnetic remanence as a function of temperature for all the hysteresis loops in the temperature scan.
Here, the MCD ratio in remanence is defined as $\frac{MCD(H = 0, \text{down sweep}) - MCD(H=0,\text{up sweep})}{2}$.
In this plot, the systematic variation of $T_C$ with the growth temperature is clear.
For the 300 $\degree$C sample, the $T_C$ is $\sim 220$ K which is consistent with the bulk $T_C$ of FGT. Increasing the growth temperature to 325 $\degree$C and 350 $\degree$C leads to increased $T_C$ of $\sim 240$ K and $\sim 260$ K, respectively. 
Finally, for the 375 $\degree$C sample, the ferromagnetism persists beyond room temperature to above 320 K, the maximum temperature of our measurement system.

\section{Discussion} \label{sec:discussion}
Based on the HAADF STEM images of the 325\degree C and 375\degree C FGT/Bi$_2$Te$_3$ samples, we learn that the higher temperature growths of FGT on Bi$_2$Te$_3$ lead to the formation of thicker vdW FGT layers with varying stoichiometry of Fe. Specifically, there is coexistence of Fe$_3$GeTe$_2$, Fe$_5$Ge$_2$Te$_2$ and Fe$_7$Ge$_6$Te$_2$ in our 375\degree C grown FGT. The most well-known composition of Fe$_m$Ge$_m$Te$_2$ is Fe$_3$GeTe$_2$, which has been reported to have a bulk T$_{C}$ of $\sim$220-230 K. 
This value is consistent with our 300\degree C and 325\degree C samples with mostly uniform vdW Fe$_3$GeTe$_2$ layers showing up in the STEM images. For the Fe$_5$Ge$_2$Te$_2$ layer, however, it is reported to have a T$_{C}$ of $\sim$250 K and possess a large PMA \cite{jothi_fe_2020}. A similar structure Fe$_5$GeTe$_2$ was also reported in literature, and it was synthesized to establish magnetic order up to $\sim$310 K for bulk materials and $\sim$280 K for thin exfoliated flakes \cite{may_ferromagnetism_2019}.
As for the thickest vdW FGT layer Fe$_7$Ge$_6$Te$_2$ that we observed in the 375\degree C sample, the magnetic properties of it are still unknown. It is possible that with the increased number of Fe atoms in one vdW layer, T$_{C}$ of this component could be even higher. 
In the literature, monolayer Fe$_7$GeTe$_2$ was calculated to have a T$_{C}$ of $\sim$570 K \cite{liu_layer-dependent_2022}. But there has been no report of measurements on Fe$_7$Ge$_6$Te$_2$ so far. 
Intercalants are also observed in the vdW gaps. 
Recent studies show that Fe intercalants in the vdW gaps of Fe$_3$GeTe$_2$ are responsible for the enhancement of $T_C$ from 160 K to 230 K in bulk crystals \cite{wu_feintercalation_2023}.
Thus, the elevated Curie temperature may be due to a combination of different thicker layers of Fe$_m$Ge$_n$Te$_2$ in the film, the interlayer coupling/proximity effect between them, and intercalants in the vdW layers.

Looking back at our temperature-dependent growths, it is rather surprising that the temperature 375\degree C could help stabilize three different layers of FGT at the same time, with each of them containing a different stoichiometry of Fe respectively. At relatively lower temperatures, the stoichiometry of Fe$_3$GeTe$_2$ seems to be the most energetically favored state. From what we have seen in the STEM images, when we use a higher growth temperature, the Fe and Ge atoms are more likely to bond together to form thicker Fe-Ge layers before the Te atoms come into play and separate them.
Thus, at higher growth temperatures, the greater mobility of the atoms enables epitaxial stabilization of different material phases of the FGT family. 

 Our findings reveal the possibility of integrating different compositions of FGT on topological insulators, with still a very sharp interface in between. At this particular temperature (of 375\degree C), all the parameters delicately balance with each other and lead to coexistence of different phases of FGT. 
 This offers a playground for interesting growths of 2D magnet FGT/TI heterostructures with a lot of tunability. 
 If we consider breaking the balance by tuning 
 growth conditions, it might become possible to stabilize targeted Fe$_m$Ge$_n$Te$_2$ compositions for tailored magnetic properties of the heterostructures. As we have studied in our previous work of kinetically controlled FGT growth on Ge substrates \cite{zhou_kinetically_2022}, the growth rate can also be a useful knob to tune, stabilize and optimize certain components of our vdW material during growth. There are also many other important parameters such as the Fe:Ge atomic ratio that could be changed accurately to help tune composition of our material. Fine tuning of the temperature could also play an important role. With these knobs, we can anticipate that tailoring the properties of the 2D magnet/TI heterostructures will become easier and more versatile.

\section{Conclusion} \label{sec:conclusion}

In summary, we demonstrate tuning of the Curie temperature to above room temperature in MBE-grown FGT/Bi$_2$Te$_3$ heterostructures by varying growth conditions. Specifically, variable-temperature magneto-optic measurements reveal that higher growth temperatures lead to higher Curie temperatures, with sample growth at 375\degree C exhibiting ferromagnetism that persists to over 320 K.
Cross-sectional STEM imaging of samples grown at 325$\degree$C and 375$\degree$C identifies differences in the atomic-scale structure.
Compared to FGT films grown at lower temperatures (325\degree C), in which we have uniform Fe$_3$GeTe$_2$ layers integrated with Bi$_2$Te$_3$, thicker vdW layers in the FGT family and intercalation (likely Fe atoms) are also observed for growth at 375\degree C.
We find that highly-ordered Fe$_5$Ge$_2$Te$_2$ and Fe$_7$Ge$_6$Te$_2$ coexist with Fe$_3$GeTe$_2$, while still keeping the interface with Bi$_2$Te$_3$ sharp and clean. 
The thicker FGT layers and intercalation that form in the higher-temperature growths both likely contribute to the enhanced Curie temperature because a number of members in FGT family with more Fe atoms in the single vdW layer have been reported and/or predicted to have higher Curie temperatures than Fe$_3$GeTe$_2$ (T$_C$ $\sim$220 K), and Fe intercalants could enhance the interlayer coupling.
These are very interesting systems for spintronics devices and for studying phenomena related to the topological surface states, especially since ferromagnetism is realized above room temperature, high-quality interfaces are preserved, and epitaxial growth enables scalability.

\section*{Acknowledgements}
\label{sec:acknowledgements}

The research at Ohio State was supported by the AFOSR/MURI project 2DMagic (FA9550-19-1-0390) and the US Department of Energy (DE-SC0016379). K.R. was supported by DAGSI (RX22-OSU-22-1). I.L. was supported by the Center for Emergent Materials, an NSF MRSEC (DMR-2011876). The research at Cornell was supported by the AFOSR/MURI project 2DMagic (FA9550-19-1-0390) and the US National Science Foundation (DMR-2104268). T.M.J.C. was supported
by the Singapore Agency for Science, Technology, and Research, and Y.K.L. acknowledges the Cornell Presidential Postdoctoral Fellowship. The work utilized the shared facilities
of the Cornell Center for Materials Research (supported by the NSF via grant DMR-1719875) and the Cornell NanoScale Facility, a member of the National Nanotechnology Coordinated
Infrastructure (supported by the NSF via grant NNCI-2025233).


\bibliography{main.bib}

\end{document}